\documentclass{article}

\usepackage{arxiv}

\usepackage[utf8]{inputenc} 
\usepackage[T1]{fontenc}    
\usepackage{hyperref}       
\usepackage{url}            
\usepackage{booktabs}       
\usepackage{amsfonts}       
\usepackage{nicefrac}       
\usepackage{microtype}      
\usepackage{makecell}
\usepackage{siunitx}
\usepackage{caption}
\usepackage{subcaption}
\usepackage{multirow}
\usepackage{xr}
\usepackage{cleveref}
\usepackage{graphicx}
\usepackage{epsfig}
\usepackage{tikz}

\pdfoutput=1

\title{Understanding (Ir)rational Herding Online}

\author{
    Henry K. Dambanemuya \\
  Northwestern University\\
  Evanston, IL 60208 \\
  \texttt{hdambane@u.northwestern.edu} \\
   \And
  Johannes Wachs \\
  Corvinus University of Budapest \\
  Budapest, Hungary \\
  \texttt{johanneswachs@gmail.com} \\
  \And
  Em\H oke-\'Agnes Horv\'at \\
  Northwestern University\\
  Evanston, IL 60208 \\
  \texttt{a-horvat@northwestern.edu } \\
}

\begin{document}
\maketitle

\begin{abstract}
Investigations of social influence in collective decision-making have become possible due to recent technologies and platforms that record interactions in far larger groups than could be studied before. Herding and its impact on decision-making are critical areas of practical interest and research study. However, despite theoretical work suggesting that it matters whether individuals choose who to imitate based on cues such as experience or whether they herd at random, there is little empirical analysis of this distinction. To demonstrate the distinction between what the literature calls ``rational'' and ``irrational'' herding, we use data on tens of thousands of loans from a well-established online peer-to-peer (p2p) lending platform. First, we employ an empirical measure of memory in complex systems to measure herding in lending. Then, we illustrate a network-based approach to visualize herding. Finally, we model the impact of herding on collective outcomes. Our study reveals that loan performance is not solely determined by whether the lenders engage in herding or not. Instead, the interplay between herding and the imitated lenders' prior success on the platform predicts loan outcomes. In short, herds led by expert lenders tend to pick loans that do not default. We discuss the implications of this under-explored aspect of herding for platform designers, borrowers, and lenders. Our study advances collective intelligence theories based on a case of high-stakes group decision-making online.
\end{abstract}


\section{Introduction}

Individuals are often exposed to others’ opinions online before forming and expressing their own. Just as we tend to follow people walking ahead of us, we often imitate others in online settings. It makes sense to do this during decision-making, as gathering exhaustive information on all available choices is typically impossible. Existing studies offer substantial evidence that the social influence that arises from prior behaviors and opinions can influence peoples’ decision-making~\cite{salganik2006experimental,muchnik2013social}. Observing others' behavior plays an increasingly dominant role in shaping individual judgment on many online platforms, such as social media sites, peer-production communities, and other peer-to-peer (p2p) platforms~\cite{kim2015online,huffaker2010dimensions,moldon2021gamification}.

Out of various potential examples, online p2p lending is an ideal setting to study collective intelligence in the presence of social influence. Like many other online platforms, it is a low-information and high-uncertainty environment. Specifically, p2p lending is characterized by a broad spectrum of borrowers, different types of loan requests, and untrained lenders relying on sparse data to assess creditworthiness. What distinguishes p2p lending from other online platforms is that it is also a high-risk setting, where lenders incur significant opportunity costs of time and monetary investment in case of loan default.

Ample research points to various factors that shape individuals' judgments on p2p lending sites~\cite{chan2020bellwether,lee2020new,stevenson2019out,vismara2018information,zaggl2019small,zhang2012rational}. Collective intelligence research suggests, for instance, that social influence can lead to people conforming to the behaviors or opinions of others even if they have private conflicting information~\cite{banerjee1992simple,haghani2019imitative}. Doing so can lead to various results in practice that range from efficient outcomes (wisdom of the crowds) to collective failure (madness of the crowds or bubbles). Existing literature demonstrates the beneficial effects of social influence on belief accuracy in estimation tasks~\cite{farrell2011social,gurcay2015power,becker2017network,navajas2018aggregated,becker2019wisdom,jansen2016forecasting}, revenue and sales forecasting~\cite{cowgill2015corporate,atanasov2017distilling,da2020harnessing}, and predicting the success of advertising campaigns~\cite{hartnett2016marketers}. At the same time, there is evidence of social influence causing decisions to converge on sub-optimal outcomes when the initial decisions of a few early movers sway subsequent choices by others in information cascades~\cite{easley2010networks,haghani2019imitative,vismara2018information}. These seemingly conflicting strands of literature leave us asking: Under which conditions does social influence enhance or diminish collective outcomes? 

Earlier work by Zhang and Liu~\cite{zhang2012rational} has formalized the beneficial vs unhelpful social influence scenarios as \emph{rational} vs \emph{irrational herding}. In particular, when herding individuals investigate alternatives, for instance, by choosing to imitate the choices of experts with a successful track record, they behave more rationally than if they simply imitate at random or by primacy. Yet while previous work discusses potential mechanisms that explain how people choose to herd rationally vs irrationally and how these decisions lead to collective consequences, the empirical validity of these theories remains relatively untested. This likely owes to the absence of a strong measurement of herding and the difficulty of procuring data on herding behavior and outcomes. Additionally, since most prior studies are based on online experiments~\cite{becker2019wisdom,becker2020network}, prediction markets~\cite{atanasov2017distilling,wolfers2004prediction,arrow2008promise}, or competitions~\cite{surowiecki2005wisdom,hong2004groups,malone2010collective}, it is unclear how existing findings on the link between herding and collective outcomes translate to real-world problems where individuals have financial and social incentives, diverse resources and constraints, and interactions are mediated by platforms that potentially change their design throughout reasonable observation periods~\cite{malik2016identifying,wagner2021measuring}. To fill this gap in the literature, in this paper, we empirically study the relationship between herding and collective outcomes in the crucial real-world setting of p2p lending by adopting a correlation-based measure of memory in complex systems to discover the conditions under which herding enhances or undermines collective decision-making. 

Our work contributes to the growing literature on herding~\cite{banerjee1992simple,bikhchandani1998learning,cai2009observational} and crowdfunding~\cite{agrawal2015crowdfunding,astebro2019herding,vismara2018information,ceyhan2011dynamics,dambanemuya2019harnessing,dambanemuya2021multi,stevenson2019out,horvat2015network,lee2020new,horvat2023hidden}. The proposed empirical framework and measure of herding are also applicable to a variety of socio-technical systems such as recommendation engines (e.g., to enhance product discovery), deliberative prediction markets (e.g., forecasts about future trends, technological innovations, or market developments), and early disaster warning and evacuation systems (e.g., escaping from building fires or natural disasters).

\section{Herding in Crowdfunding}
\label{sec:rw}

P2p lending is one of the most common forms of crowdfunding that allows borrowers to receive varying amounts of interest-based unsecured loans from the crowd~\cite{ceyhan2011dynamics,zhang2012rational,huang2021message,mild2015low,dambanemuya2019harnessing}. Borrowers typically request loans of varying amounts through project listings on dedicated online platforms. Project listings describe the characteristics of the loan, such as the amount requested, interest rate, and monthly payment. Borrowers also provide information about their credit grade, debt-to-income ratio, and whether they are homeowners. Potential lenders decide which projects to fund based on this limited information. 

In crowdfunding literature, signaling theory~\cite{spence1973job} has been widely used to explain why lenders select certain projects over others~\cite{chan2020bellwether,stevenson2019out}. According to this theory, information about borrowers and their loans visible to lenders on online p2p lending platforms signals creditworthiness. In the short term, lenders look for clues to determine which listings will likely materialize into loans. In other words, they are selecting projects that they perceive to be appealing to other lenders too, such that the project will manage to raise the target amount. In the long term, lenders' decisions also reflect their anticipated return on the investment. Lenders are searching for signals that might correlate with the likelihood that the borrower will repay the loan. 

However, determining which borrowers may default based on limited information is a non-trivial task for crowds of untrained lenders who face significant information asymmetries and social influence compared to offline lenders. P2p lenders may have limited access to comprehensive borrower information, unlike traditional lending institutions. They typically rely on the borrowers' information, such as credit scores, income statements, and loan purposes. However, borrowers may not always provide accurate or complete information, leading to information gaps and a potential asymmetry whereby one party (in this case, the lender) has less information than the other party (the borrower) in a transaction. Additional information asymmetries can also arise among lenders themselves due to their varying levels of experience and expertise in lending practices and limited mechanisms for sharing information with each other. When uncertain about whether and how to allocate their funds, online p2p lenders may learn by observing other lenders' activities that are visible on the platform~\cite{croson2008impact}.

While such social learning is necessary, especially for novice lenders~\cite{zakhlebin2019investor}, blindly imitating others can result in undesired herding. Zhang and Liu~\cite{zhang2012rational} distinguish between rational and irrational herding based on the accuracy of decision-making. In their definition, herding is rational when it improves the lending crowd's collective accuracy in selecting good from bad listings and results in successful loan repayment. Conversely, herding is irrational when lenders herd on a loan that ultimately defaults.

Herding is one of the most fundamental and widely discussed forms of social influence in crowdfunding. The literature documents a positive reinforcement effect whereby prior contributions lead to more follow-up contributions~\cite{colombo2015internal,vismara2018information,zhang2012rational,astebro2019herding,kuppuswamy2017does}. In contrast to this ``success breeds success'' perspective~\cite{van2014field}, a growing body of work demonstrates evidence for reverse herding. Specifically, small prior contributions can lead to a reduction of follow-up contributions~\cite{zaggl2019small,koning2013experimental}. A common explanation for this is that small funding amounts, relative to no contributions at all, may provide a salient signal for lenders' uncertainty or hesitation towards a project's merits, ruling out the possibility that lenders did not see the listing yet and anchoring future contributions. Recent research also proposes a ``U-shaped'' relationship between initial funding and subsequent contributions. Based on that work, the lowest subsequent contributions are associated with medium prior funding~\cite{chan2020bellwether}. 

Although this body of literature is inconclusive about the exact relationship between subsequent contribution amounts, it reinforces the fundamental observation that herding is reflected in the amounts lenders choose to contribute to the loan. Inspired by this critical insight, a crucial aspect of our current work is to develop an empirical measure of herding based on contribution amounts. This approach to quantifying herding is markedly different from existing attempts that measure herding in terms of the presence or absence of follow-up contributions~\cite{zhang2012rational,chan2020bellwether,zaggl2019small,astebro2019herding,vismara2018information}. We provide a fine-grained measure of herding based on similarities in consecutive contribution amounts such that herding is detected when lenders imitate the most recent contribution amounts. 

Though formally defined via a correlation between contribution amounts, our herding measure gains additional richness using a network perspective. Ample studies demonstrate that the communication network structure in a group mediates the relationship between social influence and collective intelligence~\cite{lazer2007network,becker2017network,barkoczi2016social}. For instance, Becker et al~\cite{becker2017network} show that the dynamics of collective decision-making change with network structure. In their experiments of numerical forecasts, even as individual beliefs become more similar (i.e., herding increases), social influence improves the accuracy of group estimates in decentralized communication networks. While most p2p lending platforms do not involve an explicit communication network between lenders, research has argued for the value of studying co-lending networks to trace which lenders tend to contribute to the same projects repeatedly~\cite{horvat2015network}. This paper introduces a novel approach to visualize herding via lenders contributing similarly to the \emph{same} loan.

Finally, our research also draws on prior work that links lenders' self-efficacy to their decision-making performance~\cite{stevenson2019out}. Existing work on p2p lending quantifies self-efficacy as \emph{lenders' prior success} in selecting repaid listings~\cite{dambanemuya2019harnessing}. Typically, crowdfunding platforms comprise a variety of lenders ranging from novices to expert lenders. However, existing studies show that novice lenders are less likely to recognize and react to negative pitch cues when there are positive signals from the crowd, leading them to invest nearly three times as much in a poor-quality venture than expert lenders~\cite{stevenson2019out}. At the same time, collective intelligence theory suggests that diverse opinions are more likely to lead to superior outcomes, even in the absence of experts~\cite{page2008difference,malone2010collective}. Thus, this work links lenders' self-efficacy to herding and investigates how herding among lenders with good vs bad track records impacts collective outcomes.

\section{The Prosper Marketplace}
We use data from the Prosper marketplace, one of the oldest p2p lending platforms in the US. The platform acts as a broker between borrowers seeking loans outside traditional financial institutions and individual lenders who contribute small amounts towards the requested sum. The lending process is characterized by competition between borrowers,  individual lenders' decision-making, and herding between lenders. Competition arises as multiple listings are available simultaneously to each lender. Lenders have the autonomy to choose from these listings. Based on the observed partial information about the characteristics of the listing and borrower, lenders must decide whether to lend and, if so, how much. They can observe others' contributions and monitor the evolution of the loan~\cite{burtch2013empirical,ceyhan2011dynamics}, engaging in social learning and potentially rational or irrational herding. When a listing reaches its target amount, the lenders' contributions are pooled into a single loan awarded to the borrower at a final interest rate (between $6.95\%$ and $35.99\%$ for first-time borrowers) determined by several factors, including the borrower's credit history, the amount requested and the payment period for the loan. In the case of listings that fail to raise the target amount, lenders are refunded all their contributions.

\begin{figure*}[!h]
    \centering
    \includegraphics[scale=.4]{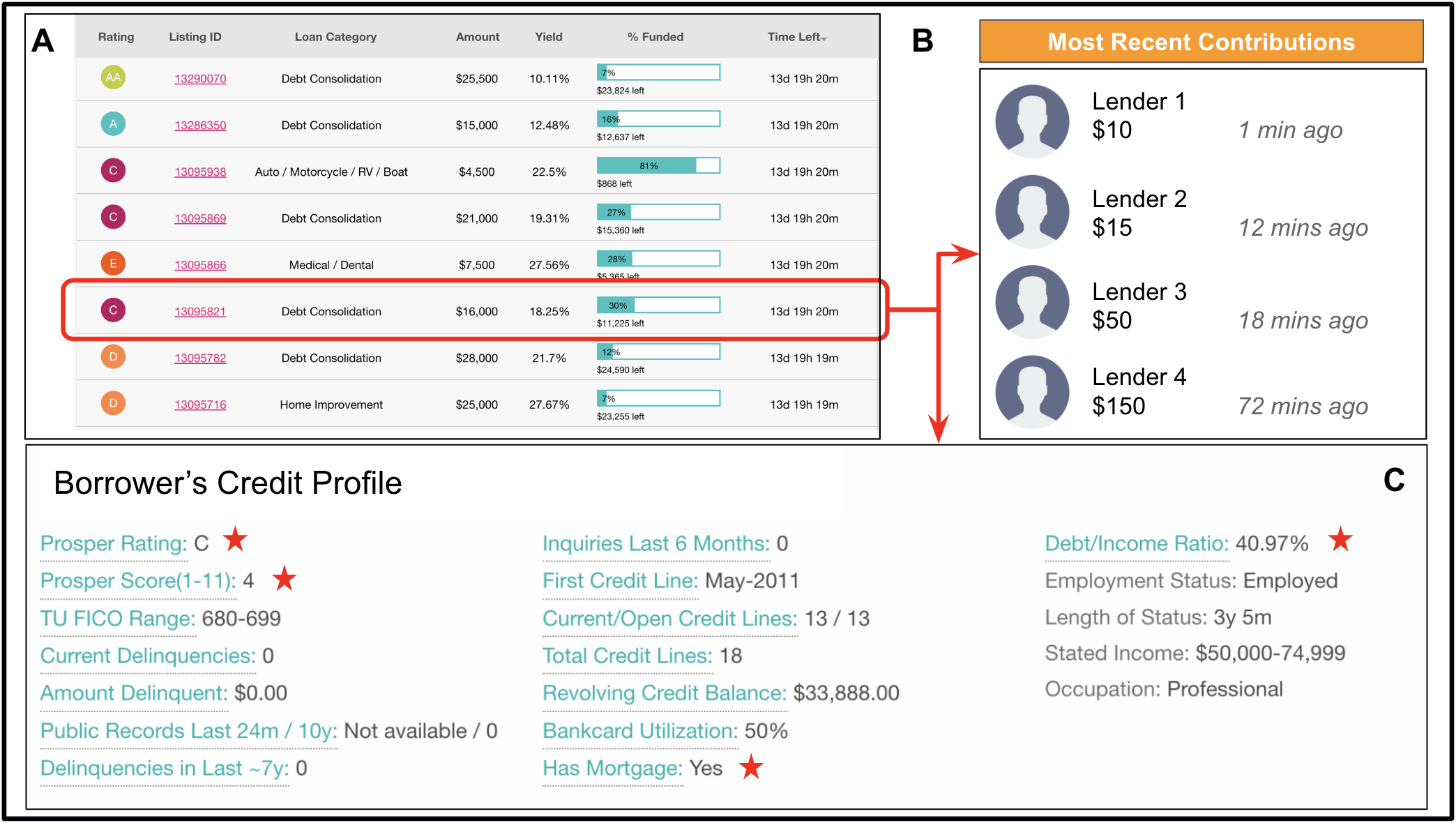}
    \caption{The p2p lending platform's user interface shows (a) different project listings, (b) the most recent contributions to one of the listings, and (c) the borrower's credit profile. Key variables obtained from the borrower's profile are indicated with red stars. The detailed description associated with the loan listing is not shown here.}
    \label{fig:mockup}
\end{figure*}

\paragraph{Loan Information:} The data comprise $27,624$ funded loan listings created between November $2005$ and October $2008$. The listings can belong to one of six loan categories: i.e., auto, personal, business, student, home improvement, or debt consolidation loans (Figure~\ref{fig:mockup}A). The loans were funded from $3,556,673$ contributions made by $51,612$ unique lenders and amounted to $\$174,619,263$. Borrowers can request loans between $\$1,000$ and $\$25,000$. Our data is comprised of listings with an average of $\$6,321$ in \emph{amount requested}. The listings are active for a maximum of $14$ days. For each loan listing, borrowers must include a written statement that describes the purpose of the loan. The average length of project descriptions in our data is $224$ words. Our \emph{description length} variable is measured in hundreds of words. 

\paragraph{Borrower Profiles:} On the Prosper Marketplace, borrowers are required to provide credit information (Figure~\ref{fig:mockup}C). This includes their \emph{homeownership} status (yes/no). In our data, $44.45\%$ of the borrowers own their homes. The credit profile also includes the \emph{debt-to-income ratio}, calculated as monthly debt payments divided by monthly income. On average, the borrowers have a debt-to-income ratio of $0.320$. Borrowers are assigned a \emph{credit grade} based on their creditworthiness. The credit grade ranges from 2 (HR = high risk) to 8 (AA = best credit). We excluded borrowers with no credit. The median credit grade of all borrowers is C. Additionally, borrowers are assessed using the \emph{Prosper score}, a custom risk score built using historical Prosper data on the borrower's risk level. The Prosper score ranges from 1 to 11, with 11 being the lowest risk and 1 being the highest risk. The most common Prosper score is $1$, accounting for $13.88\%$ of the borrowers. The default rate changes from 15.4\% for the listings with the best credit to 61.8\% for those with the worst credit. Taking a risk on borrowers with bad credit might be attractive to lenders due to higher interest rates.

\paragraph{Lending Dynamics:} Ample research on multiple crowdfunding platforms shows that characteristics of the lending dynamics can be as or even more predictive of collective outcomes than loan and borrower information~\cite{dambanemuya2019harnessing,dambanemuya2021multi,horvat2023hidden}. Using various computational methods, these studies find that variations in the timings and amounts of lenders' contributions are positively associated with both loan funding and payment success. Thus, in our analysis, we include control variables that describe the lending dynamics that characterize the frequency of contributions, the speed at which the funds are accumulating, and the opinion diversity reflected in individual contribution amounts. 

Specifically, we measure the \emph{number of contributions} for each listing and expect this to correlate with success~\cite{dambanemuya2021multi,ceyhan2011dynamics}. The average listing receives $134$ contributions with a $\$79.57$ mean contribution amount. 

We also consider the main temporal aspect of lenders' contribution activity by measuring the \emph{momentum} with which contributions arrive. The momentum is quantified by the ratio between the mean and standard deviation of the times between consecutive contributions~\cite{horvat2023hidden,dambanemuya2021multi}. We expect this measure to signal lenders' confidence in a project's merits. For example, consider a campaign with contributions that arrive $1$, $12$, $18$, and $72$ minutes after the campaign's launch (c.f. Figure~\ref{fig:mockup}B). The corresponding inter-contribution times are $1$, $11$, $6$, and $54$ minutes, which yields a momentum of $1.35$. 

For each listing, we further quantify the \emph{opinion diversity} reflected in the contribution amounts as the ratio between the mean and standard deviation of the contribution amounts~\cite{dambanemuya2021multi}. For example, a campaign receiving the amounts \$10, \$15, \$50, and \$150 has an opinion diversity of $1.16$ (c.f. Figure~\ref{fig:mockup}B). Slightly different operationalizations of momentum and opinion diversity lead to similar results~\cite{horvat2023hidden}. 

\paragraph{Lender Self-Efficacy:} Lenders on the Prosper marketplace comprise individuals with varying levels of experience in investing. To investigate whether and how lender experience influences collective outcomes, we incorporate a measure of lenders' success on the platform~\cite{dambanemuya2019harnessing}, quantified as the fraction of successfully paid loans over all the loans that a lender contributed to up the time of the current decision. The lenders' previous success rate ranges from 0 to 1 with a median of 0.64 in our data. Previous success is a measure at the level of individual lenders and changes as we re-calculate it after each project listing the lender contributed to based on the outcome of the loan (repaid or defaulted). Note that some of our main analyses are at the level of listings, where we \emph{average over lenders' prior success rate} to describe the overall track record of lenders involved in a loan. While this is a simplification, our analyses show that it still captures the main signal related to lender expertise.

\begin{figure*}[!h]
    \centering
    \includegraphics[scale=.375]{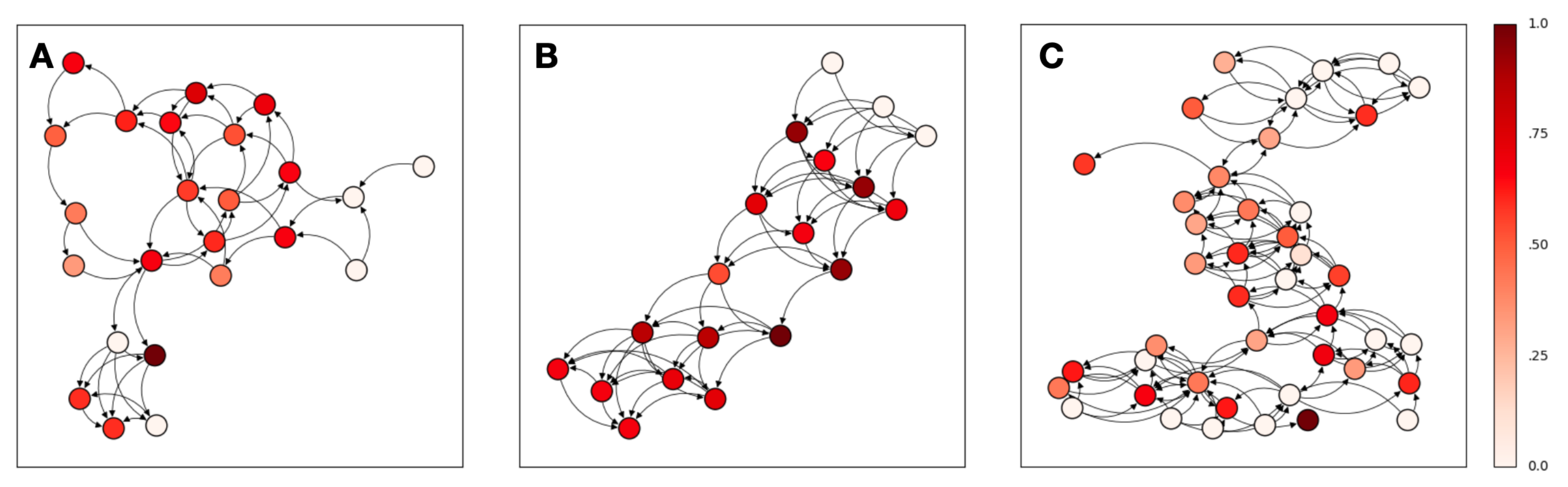}
    \caption{Examples of herding networks: (A) No herding on a defaulted loan: negative coefficient of herding (CoH) and several lenders with a poor track record; (B) Rational herding on a paid loan: positive CoH and herding follows lenders with a good track record; and (C) Irrational herding on a defaulted loan: positive CoH but lenders with low prior success are imitated. Node color represents the lenders' previous success in identifying on the platform subsequently repaid loans. Success is measured here at individual lenders' level and is expressed as the fraction of prior listings the lender contributed to and did not default.}
    \label{fig:herding-illustration}
\end{figure*}

\section{Quantifying Herding}

We quantify herding at the level of individual project listings. We measure the degree of herding on each project based on auto-correlations between lenders' consecutive contribution amounts. As discussed in Section~\ref{sec:rw}, this quantification improves on prior approaches that identify herding via the presence of follow-up contributions by drawing on literature on the importance of contribution amounts~\cite{horvat2023hidden,dambanemuya2021multi,dambanemuya2019harnessing,zaggl2019small,colombo2015internal}. For our herding measure to capture imitation in a reasonable time frame, we evaluate correlations between contribution amounts when observational learning is likely, that is, only for a certain number of consecutive contributions. Formally, we use a measure that is similar to a popular coefficient of memory in complex systems~\cite{goh2008burstiness}. Given the consecutive contribution amounts $\{A_1, A_2, ..., A_{N}\}$, we define the \emph{coefficient of herding} (CoH) as the ratio between the co-variation of the sequences $S1=\{A_1, A_2, ..., A_{N-m}\}$ to $Sm=\{A_m, A_{m+1}, ..., A_{N}\}$ and the product of the standard deviation of these sequences:

\begin{equation}
    \frac{1}{N} \sum_{i=1}^{N} \frac{(A_{i}-\mu_{S1}) \ ...\  (A_{i+m-1}-\mu_{Sm}) }{\sigma_{S1} \ ... \ \sigma_{Sm}}
\end{equation}

\noindent where $N$ is the number of contributions in the loan, $m$ is the memory range that specifies the number of consecutive sequences to consider, and $\mu_{S1}, \mu_{Sm}$ and $\sigma_{S1}, \sigma_{Sm}$ are the mean and standard deviation of the first ($S1$) and last ($Sm$) sequence, respectively. The coefficient ranges from $-1$ (perfect anti-herding) to $1$ (perfect herding). The observed coefficient of herding values in our data are between $-0.2$ and $0.8$ with a median value of $-0.005$. We use a memory range of $m=5$ to reflect the number of previous contributions that potential lenders can see on the default listing page. Our findings are robust to different memory ranges ($m=3$ or $m=7$).

The coefficient of herding captures economically the overall tendency to imitate contributions on a listing. However, it is oblivious to who is imitating whom. Obtaining a fuller picture of the structure of imitation relationships in conjunction with key characteristics of the lenders promises to reveal critical insights about the so far mysterious link between herding and collective outcomes. 

\paragraph{Visualizing Herding Networks.} To better characterize herding behavior on a listing, we capture the relationships between lenders based on mimicry of contributions via co-lending networks~\cite{horvat2015network}. A co-lending network is a directed graph whose nodes represent unique lenders and whose edges connect lenders with the same contribution amount within a memory range of $m=5$ consecutive contributions on the \emph{same} project. While the memory range here is motivated by the specifics of the application (i.e. the default number of previous contributions that lenders can observe on the platform), the main characteristics of the resulting networks are robust to other choices of $m$. We consider the co-lending networks unweighted, meaning that we ignore the rare cases when the same lender is imitating the same other lender multiple times on the same listing. 

Figure~\ref{fig:herding-illustration} shows example herding networks created with this process. Each node is colored based on prior lender success. Prior success represents the ratio of successful loans to all the loans a lender has contributed to. Figure~\ref{fig:herding-illustration}A illustrates a listing on which lenders did not herd. We observe that several lenders had low prior success in identifying subsequently repaid loans. This loan defaulted. Figure~\ref{fig:herding-illustration}B depicts a listing with substantial herding. More importantly, the herd is led by lenders with a strong track record. The loan is repaid, making this an exemplary case of rational herding. Finally, Figure~\ref{fig:herding-illustration}C shows another case of high herding, but here lenders with a poor track record are imitated, leading the herd to a loan that did not get repaid. These examples are representative of many other listings we inspected manually. They give us invaluable intuition about the role of the interplay between herding and lender prior success.

Note that while these network visualizations make the connection between herding, lender track record, and loan repayment almost straightforward, actors involved in p2p lending only have a local view of the system. The holistic picture we visualize at the level of listings is critical to understand the emergence of collective intelligence in p2p lending. 

\subsection{Herding in the Prosper Marketplace}
We observe that our empirical CoH is significantly and positively correlated with certain characteristics of loans, borrowers, and lending dynamics.

\paragraph{Herding and Loan Information.} The CoH is significantly positively correlated with the requested loan amount, suggesting that there is systematically more herding on large loans than smaller ones (Table~\ref{tab:corr-loan}). This is essential to keep in mind because larger fundraising goals are associated with higher default risk~\cite{zhang2012rational}. If larger target amounts also amplify herding, platform maintainers and lenders should know about it. We find no significant correlation between herding and the project description length.

\begin{table}[!h]
\centering
\begin{tabular}{lcc}
\hline
 & Requested Amount & Description Length \\ \hline
CoH & 0.178       & 0.002  \\
    & p<0.001   & p=0.684 \\ \hline
\end{tabular}%
\caption{Pearson correlations between herding and loan information.}
\label{tab:corr-loan}
\end{table}

\paragraph{Herding and Borrower Profiles.} Borrowers' creditworthiness indicators, such as homeownership, debt-to-income ratio, Prosper score, and credit grade are associated with herding to a lesser extent (Table~\ref{tab:corr-borrower}). The weak connections between established indicators of creditworthiness and herding are particularly noteworthy, as they suggest strong concerted interest in potentially risky project listings. P2p lending offers opportunities to borrowers excluded from traditional financing due to poor credit. However, these borrowers can still represent good investments due to higher interest rates, and p2p lenders are willing to take a chance, herding on such listings. 

\begin{table}[!h]
\centering
\begin{tabular}{lcccc}
\hline
 & \begin{tabular}[c]{@{}l@{}}Home- \\ Ownership\end{tabular} & \begin{tabular}[c]{@{}c@{}}Debt-to-Income \\ Ratio\end{tabular} & \begin{tabular}[c]{@{}l@{}}Credit \\ Grade\end{tabular} & \begin{tabular}[c]{@{}l@{}}Prosper \\ Score\end{tabular} \\ \hline
CoH  & 0.032    & 0.013   & 0.097    & 0.092  \\
    & p<0.001  & p=0.015 & p<0.001  & p<0.001\\ \hline
\end{tabular}%
\caption{Pearson correlations between herding and borrower profiles.}
\label{tab:corr-borrower}
\end{table}

\paragraph{Herding and Lending Dynamics.} As expected, the herding coefficient is also significantly correlated with characteristics of the lending dynamics (Table~\ref{tab:corr-dynamics}). We observe the strongest positive correlation between the number of contributions and herding. Our measure of herding does not explicitly normalize with the contribution count. This relatively high correlation arises because the involvement of more lenders allows for more imitation. However, the correlation is far from perfect ($\rho=0.250$), suggesting that the CoH contains valuable additional information compared to the sheer number of contributions. 

Herding is also significantly and positively correlated with variations in the inter-contribution times or momentum. This makes sense, as we expect high herding to shorten and homogenize the inter-contribution times, leading to a higher momentum. 

Next, the more herding there is, the less variation we observe in the contribution amounts. In other words, higher herding is associated with lower opinion diversity, leading to a low, but significant negative correlation ($\rho=-0.051$). 

Finally, we find that herding is negatively correlated with lenders' prior success. This means that herding is high in the presence of lenders with a bad track record, while more experienced lenders are less likely to herd. This observation also has critical implications for lenders and p2p platforms. In conjunction with our result that herding is high on risky project listings, the finding that it also draws more inexperienced lenders, stresses the importance of properly monitoring and addressing herding in online p2p lending.

\begin{table}[!h]
\centering
\begin{tabular}{lcccc}
\hline
 & Number of & Momentum & Opinion & Avg. Prior \\ 
 & Contributions & & Diversity & Lender Success\\ \hline
CoH & 0.250    & 0.143  & -0.051    & -0.038 \\
    & p<0.001  & p<0.001 & p<0.001  & p<0.001 \\ \hline
\end{tabular}%
\caption{Pearson correlations between herding and lending dynamics.}
\label{tab:corr-dynamics}
\end{table}

\section{Herding and Loan Repayment}

To test the relationship between herding and loan repayment outcomes, we fit a logistic regression model:

\begin{equation}
    r_{i} = \beta_1 CoH + \beta_2 AvgLenderPriorSuccess + \beta_3 (CoH\times AvgLenderPriorSuccess) + X_{i}\beta + \lambda_{i}+\delta_{i} + \eta_{i} + \epsilon
\end{equation}

\noindent where $r_{i}$ is the binary outcome corresponding to whether loan $i$ was repaid (1) or not (0), $CoH$ is the coefficient of herding, $AvgLenderPriorSuccess$ is the average previous success rate of lenders participating in the loan, $X_{i}$ is a matrix of control variables, $\lambda_{i}$ and $\delta_{i}$ are the year and category fixed effects, and $\eta_{i}$ is a dummy variable taking the value of 1 if the loan recipient is a homeowner. The control variables include the amount requested, the number of contributions received, the momentum of contributions, the opinion diversity reflected in contribution amounts, the length of the project description, the borrower's debt-to-income ratio, Prosper score, and credit grade. In addition to the full model, we fit three intermediate models: a controls-only model excluding both the herding and lender success variable, and two models containing only one of the two key variables. We show all four models in Table \ref{tab:regressions}, reporting heteroskedasticity robust standard errors for all estimates.

\paragraph{Baseline Model}

The control variables behave as expected (model 1). Campaigns receiving more contributions are more likely to be repaid. Loans asking for less money are more likely to be repaid. The higher the borrower's Prosper score the greater the chance that the loan will be repaid. Signals of external economic health (i.e., if the borrower has a low debt-to-income ratio and a high credit grade) are also significant positive predictors of loan repayment. Finally, of the two variables quantifying lending dynamics, opinion diversity has a positive significant association with loan repayment, while momentum has no significant relationship with the outcome. The year and category fixed effects included in our model mean that these relationships are robust to overall time-varying trends on the platform and idiosyncrasies between the types of fundraising efforts hosted on the site. We also observe significant stability of the control variable coefficient estimates across all four models.

\begin{table}[t]
\resizebox{\columnwidth}{!}{%
\def\arraystretch{0.70}
\begin{tabular}{lcccc}
\tabularnewline\midrule\midrule
Dependent Variable & \multicolumn{4}{c}{Loan Repaid (1) or Defaulted (0)}\\
 & (1) & (2) & (3) & (4)\\
\midrule \emph{Variables} &   &   &   &  \\
log(Number of Contributions) & 0.315$^{***}$ & 0.266$^{***}$ & 0.310$^{***}$ & 0.264$^{***}$\\
  & (0.026) & (0.026) & (0.026) & (0.027)\\
Momentum & -0.008 & -0.017 & -0.009 & -0.016\\
  & (0.021) & (0.021) & (0.021) & (0.021)\\
Opinion Diversity & 0.178$^{***}$ & 0.177$^{***}$ & 0.177$^{***}$ & 0.178$^{***}$\\
  & (0.027) & (0.027) & (0.027) & (0.027)\\
log(Amount Requested) & -0.714$^{***}$ & -0.623$^{***}$ & -0.714$^{***}$ & -0.623$^{***}$\\
  & (0.034) & (0.034) & (0.034) & (0.034)\\
Prosper Score & 0.155$^{***}$ & 0.134$^{***}$ & 0.155$^{***}$ & 0.133$^{***}$\\
  & (0.006) & (0.007) & (0.006) & (0.007)\\
log(Debt-to-Income Ratio+0.01) & -0.172$^{***}$ & -0.137$^{***}$ & -0.173$^{***}$ & -0.136$^{***}$\\
  & (0.018) & (0.019) & (0.018) & (0.019)\\
Credit Grade & 0.393$^{***}$ & 0.302$^{***}$ & 0.393$^{***}$ & 0.301$^{***}$\\
  & (0.011) & (0.012) & (0.011) & (0.012)\\
Description Length & 0.015 & 0.022$^{*}$ & 0.015 & 0.023$^{*}$\\
  & (0.010) & (0.010) & (0.010) & (0.101)\\
Avg. Prior Lender Success &    & 3.62$^{***}$ &    & 3.71$^{***}$\\
  &    & (0.229) &    & (0.228)\\
CoH &    &    & 0.184 & -3.45$^{*}$\\
  &    &    & (0.207) & (1.40)\\
Avg. Prior Lender Success $\times$ CoH &    &    &    & 5.96$^{*}$\\
  &    &    &    & (2.32)\\
\midrule \emph{Fixed-effects} &   &   &   &  \\
Project Year & Yes & Yes & Yes & Yes\\
Category & Yes & Yes & Yes & Yes\\
Homeowner & Yes & Yes & Yes & Yes\\
\midrule \emph{Fit statistics} &   &   &   &  \\
Observations & 27,344 & 27,344 & 27,344 & 27,344\\
Pseudo R$^2$ & 0.119 & 0.127 & 0.119& 0.127\\
BIC & 31,810.5 & 31,539.1 & 31,810.9& 31,552.0\\
\end{tabular}
}
\caption{Logistic regression results for control variables (model 1), average lender prior success (model 2), herding (model 3), and the interaction between lender success and herding (model 4) with loan repayment/default as the outcome variable. Heteroskedasticity-robust SE reported; Significance at p-values: \emph{***: 0.001, **: 0.01, *: 0.05}}
\label{tab:regressions}
\end{table}

\begin{figure}[!h]
    \includegraphics[scale=.9]{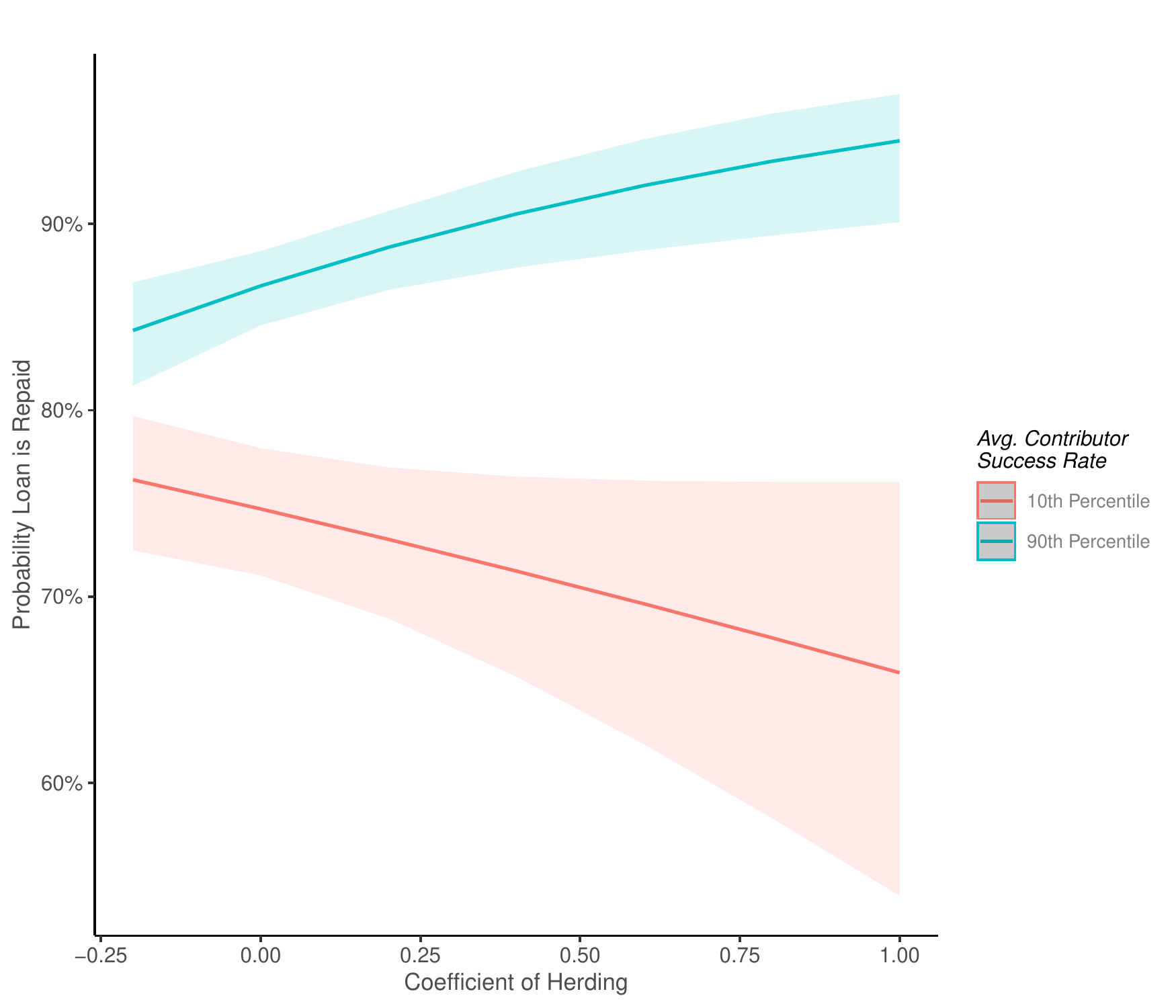}
    \caption{Model predicted chance of loan repayment as a function of the coefficient of herding (CoH). Fixing the average lender success rates in previous campaigns at the 10th and 90th percentile, respectively, highlights the significant interaction effect measured in the model. When investors with a poor previous track record herd (10th percentile), repayment is less likely, while if investors with a strong previous record herd (90th percentile), repayment is more likely.}
    \label{fig:interaction}
\end{figure}

\paragraph{Full Model} As model 4 shows, the interaction between average lenders' previous success and herding is positive and significant. If previously successful lenders are contributing to a project listing with significant herding, the loan is more likely to be successfully repaid. However, if investors with a poor prior track record are participating in herding, the loan is more likely to default. To visualize this finding, we plot the model predicted likelihood of loan repayment as a function of herding with investor previous success fixed at the 10th and 90th percentiles (Figure \ref{fig:interaction}). As noted, our findings are robust when the memory on the coefficient of herding measure is set to $m=3$ (interaction effect $3.46$, $p<.1$) or $m=7$ (interaction effect $3.80$, $p<.01$). 

All in all, tracking the presence of experienced lenders in herding networks allows distinguishing between good and bad collective outcomes (c.f. Figure~\ref{fig:herding-illustration}). Our findings suggest that the presence of experienced lenders in the herding networks is an important predictor of whether herding is rational or irrational. We find that herding around previously successful lenders results in better collective outcomes (i.e., successful loan payments) compared to when the herd follows lenders with a poor track record.

\paragraph{Intermediate Models}
The two intermediate models (2 \& 3) underscore the finding of our full model that herding is beneficial conditional on the participants having previous success. The second model, excluding herding, shows that previous success itself is a strong predictor of loan repayment. The third model includes herding but not the prior success of lenders. Here we observe that herding alone has no significant relationship with loan outcomes. 
In other words, herding only seems to matter for predicting repayment as a function of the crowd's track record and expertise. 

\section{Discussion}
Crowdfunding is a novel way of funding new ventures, enhancing entrepreneurship, and fostering innovation~\cite{mollick2014dynamics,wachs2021does}. As a form of crowdfunding, online p2p lending is a particularly important market with substantial crowd decision-making that stimulates economic growth and continues to expand, especially in sectors under-serviced by lending institutions~\cite{freedman2008social,iyer2016screening,lin2013judging,zhang2012rational,mollick2014dynamics,lee2020new}. Given the evidence that crowdfunding often acts as a seed for ideas that later attract more formal investments from venture capital \cite{kaminski2019new}, it is important to understand how the unique behavioral aspects of crowdfunding like herding and imitation among contributors can lead to virtuous or ruinous outcomes. Thus, considering the importance and high risks associated with p2p lending and that lenders on these platforms are often untrained compared to financial experts, we sought to characterize, understand, and detect instances in which lending behavior might be driven by irrational herding and hence lead to undesired lending outcomes like loan default. 

Our work draws on the conceptual framework of Zhang and Liu \cite{zhang2012rational}, which distinguishes between two possible underlying mechanisms behind an observed herd. A herd can either consist of a series of simple imitations, which Zhang and Liu call irrational, or of a series of learning events, which they call rational. In the latter case, individuals observe some information in the behavior of others and use this information to make better decisions. While both behaviors represent mental shortcuts for an investment decision, rational herding incorporates valuable information. In our context, lenders engage in rational herding when they imitate lenders with strong track records. On the platform we study, lenders could not only observe the recent contributions to a listing, but they could also observe information about users making those contributions. Clicking on a username brings a prospective lender to a user profile page with that individual's lending history on the platform. 

Consistent with signaling theory~\cite{spence1973job}, in such settings, social signaling may therefore help to overcome challenges associated with incomplete and asymmetrically distributed information between lenders and borrowers on these platforms. However, the mere presence of social signaling in the form of social influence or herding does not necessarily entail positive outcomes as the above findings show. In this work, we show how individuals participating in a p2p lending market are significantly more successful when herding rationally, that is when they imitate investors with a strong track record. In contrast, we also observe that irrational herding leads to bad outcomes in the form of loan defaults. 

\subsection{Main Findings and Contributions}

Our main finding is the empirical observation that herding behavior can lead to positive or negative outcomes relative to a campaign funded without herding, depending on whether the herd is following strong experts or random individuals. We make this contribution by analyzing a unique data set of crowdfunded loans in which we can observe the timing and size of lender contributions, as well as the history of all lenders. Crucially, prior contributions and the track record of the lenders can be accessed by prospective lenders at the time of decision. As a dependent variable, we consider not the outcome of fundraising success, as is commonly done in the crowdfunding literature, but rather the repayment of the loan to the investors, as this is the outcome of interest to the actors we study. In this way, we study success as the delivery and repayment of an investment.

With this data in hand, we develop a measure of herding in sequential decision-making that we call the coefficient of herding (CoH). The CoH describes herding in terms of correlations between consecutive contributions of a group of lenders. Similar approaches have been used in a variety of contexts, for example, to characterize memory in complex systems~\cite{goh2008burstiness}. To unpack this measure conceptually, we develop a method to visualize herding in project listings as networks in which nodes are lenders and edges denote imitation of the contribution amount. Visualizing herding networks for listings with both high and low CoH reveals that the measure captures multiple macroscopic aspects of herding in a single dimension. For instance, many smaller herds, in which small groups of individuals imitate one another and single large herds, in which everyone is imitating the same people both yield high CoH scores.

We further observe that when those who imitate actively learn or observe by following the decisions of experienced lenders with a strong record of successful investments, even novice lenders can identify which borrowers are the best lending opportunities. On the contrary, we observe irrational herding when lenders passively or reflexively mimic inexperienced lenders. As our analysis shows, such irrational herding is more likely to result in misallocating funds from meritorious projects to borrowers that default. These findings have several implications for platform creators, lenders, and borrowers.

\subsection{Implications for Platform Creators, Borrowers, and Lenders}

Quantifying herding can contribute to the \emph{stability} of the p2p lending platform. This is because herding can lead to market distortions, bubbles, or sudden shifts in lending patterns, which may undermine the platform's robustness. By actively monitoring herding, platforms can therefore identify and address these issues promptly, thereby promoting a more sustainable and balanced lending ecosystem.

Additionally, knowledge of herding can serve as a \emph{quality control} mechanism for p2p lending platforms. For example, by identifying situations where lenders may be blindly following the actions of others without conducting independent evaluations, platforms may intervene, e.g., by nudging and encouraging lenders to make independent decisions and conduct thorough assessments of borrowers. These efforts will contribute to a healthier lending ecosystem, attracting more lenders and borrowers to the platform.

For lenders, a method to quantify and visualize herding can serve as a \emph{decision support} tool. By providing quantitative and visual indicators of herding, lenders can use this information to supplement their decision-making process. For example, lenders can assess the extent of herding and the expertise of the herd better, and thus evaluate the risk associated with their lending decisions with more confidence. Hence, herding information provides valuable decision support which can ultimately lead to improved collective outcomes.

Demonstrating the ability to quantify herding behavior can also enhance lenders' \emph{confidence and trust} in p2p platforms. For example, lenders may feel more comfortable participating in a platform that actively monitors and reports herding behavior. Another significant benefit to lenders is \emph{risk mitigation}. If platforms can identify situations where herding may lead to increased default rates or systemic risks, they can proactively protect lenders from potential adverse effects caused by irrational herding, e.g., through initiatives to diversify lenders on campaigns or lender education programs.

For borrowers, a direct benefit of the above implications for lenders is also an enhanced borrower experience and \emph{better opportunities} in the p2p lending market. For example, by incorporating herding information, platforms can better assess borrowers' creditworthiness beyond traditional credit scores which can potentially lead to more favorable loan terms and improved access to loans for borrowers who may have been overlooked based solely on conventional credit metrics. 

Finally, our work facilitates ongoing research and development efforts within p2p lending platforms. We believe that by quantifying herding behavior, platforms can generate valuable data and insights that can be used for further analysis, research, and refinement of lending models and algorithms. This continuous improvement can result in more effective risk assessment methodologies, enhanced platform features, and improved overall performance.

\subsection{Limitations and Future Work} 
Our work uses data from 2005 to 2008 and the Web has changed significantly since then. To the best of our knowledge, this is the only data set available for research that contains verified information about loan repayment, which is needed to evaluate collective intelligence. Provided new data, future work could test the temporal validity of our findings taking into account the effect of platform and ecological changes.

Future work should also investigate to what extent the socio-technical design of the platforms (e.g., site personalization, gamification elements, ranking systems, and recommendation engines) might influence herding behavior. Further inquiry along these lines would help in disentangling whether herding in the web context is more a ``people problem'' or a ``platform problem.'' Future work could also investigate the effect of loan descriptions on herding. While our models include the length of the descriptions, it is worth exploring the effect of linguistic features obtained through a more advanced NLP model (e.g., transformer models) on herding. Finally, we study herding behavior in a digital platform. Thus, it remains unknown whether and how our methods and findings may translate to offline behaviors.

\section{Conclusion}
This study presented empirical findings indicating that participants in crowdfunding projects frequently imitate each other's contributions, leading to the formation of herds. However, the impact of such behavior on collective outcomes exhibited a non-uniform pattern, which is contrary to initial expectations. Specifically, herds led by expert individuals tended to select successful projects, whereas herds led by individuals with a worse track record were more prone to selecting unsuccessful projects. Within the context of our data set obtained from a leading p2p lending platform, where individual lenders provide financial support to borrowers charging interest, these findings suggest potential strategies for borrowers and lenders to enhance their returns and mitigate risks. In a broader sense, our research demonstrates that the consequences of collective imitation or herding can vary substantially depending on the composition of the herd.

\section*{Acknowledgments}
This work was supported by the U.S. National Science Foundation under Grant No. IIS-1755873.

\bibliographystyle{unsrt}  
\bibliography{references}  

\end{document}